\documentclass[11pt]{article}
\usepackage{jheppub1}
\usepackage{amsmath,amssymb,amsfonts,graphicx}

\newcommand{\be}{\begin{equation}}
\newcommand{\ee}{\end{equation}}
\newcommand{\bea}{\begin{eqnarray}}
\newcommand{\eea}{\end{eqnarray}}
\newcommand{\beas}{\begin{eqnarray*}}
\newcommand{\eeas}{\end{eqnarray*}}
\newcommand{\ba}{\begin{array}}
\newcommand{\ea}{\end{array}}
\newcommand{\tr}{{\rm tr}}

\renewcommand*\d[2][]{%
	\mathrm{d}%
	\ifx\relax#1\relax\else
	\rule{-0.02em}{1.5ex}^{#1}\rule{0.08em}{0ex}\!
	\fi
	#2\,
}

\title{Comments on wormholes, ensembles, and cosmology}

\author[]{Mark Van Raamsdonk}

\affiliation[]{Department of Physics and Astronomy, University of British Columbia,\\
6224 Agricultural Road, Vancouver, B.C.\ V6T 1Z1, Canada.}

\emailAdd{mav@phas.ubc.ca}

\abstract{Certain closed-universe big-bang/big-crunch cosmological spacetimes may be obtained by analytic continuation from  asymptotically AdS  Euclidean wormholes, as emphasized by Maldacena and Maoz. We investigate how these Euclidean wormhole spacetimes and their associated cosmological physics might be described within the context of AdS/CFT. We point out that a holographic model for cosmology proposed recently in arXiv:1810.10601 can be understood as a specific example of this picture. Based on this example, we suggest key features that should be present in more general examples of this approach to cosmology. The basic picture is that we start with two non-interacting copies of a Euclidean holographic CFT associated with the asymptotic regions of the Euclidean wormhole and couple these to auxiliary degrees of freedom such that the original theories interact strongly in the IR but softly in the UV. The partition function for the full theory with the auxiliary degrees of freedom can be viewed as a product of partition functions for the original theories averaged over an ensemble of possible sources. The Lorentzian cosmological spacetime is encoded in a wavefunction of the universe that lives in the Hilbert space of the auxiliary degrees of freedom.}

\notoc

\begin{document}

\maketitle
\newpage
\parskip=10pt

\section{Introduction}

Despite remarkable progress over the recent decades in our understanding of quantum gravity via the AdS/CFT correspondence \cite{Maldacena:1997re}, a non-perturbative description of gravitational physics in cosmological spacetimes remains elusive. Describing the physics of a closed universe seems particularly challenging, since an asymptotic boundary of space appears to play a key role in the well-understood holographically-described quantum gravitational systems.\footnote{For various other approaches to describing cosmological phyics using holography, see \cite{Strominger:2001pn, Banks:2001px,Hertog:2004rz, Alishahiha:2004md, Freivogel:2005qh, McFadden:2009fg, Banks:2018ypk}.}

\begin{figure}
\centering
\includegraphics[width=80mm]{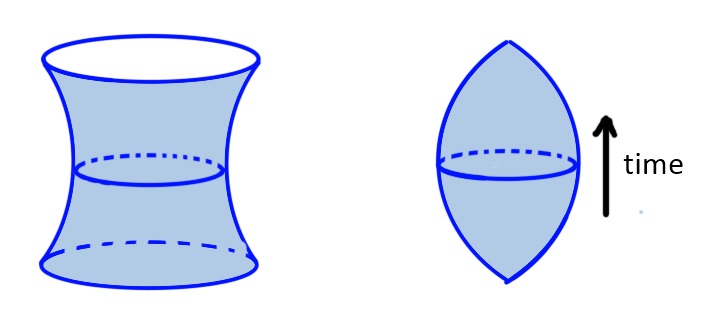}
\caption{The Maldacena-Maoz construction. Left: Euclidean AdS wormhole. Right: The analytic continuation to a closed universe big-bang / big-crunch cosmology.}
\label{fig:MMcosmo}
\end{figure}

Certain cosmological spacetimes have a Euclidean continuation with a spatial asymptotic boundary. Thus, it is possible that a theoretical description of the cosmological physics in these spacetimes could be encoded in the analytic structure of observables in some Euclidean field theory living on this boundary. A specific example of this was discussed by Maldacena and Maoz in \cite{Maldacena:2004rf}, which points out that certain analytic continuations of two-boundary asymptotically AdS Euclidean wormhole geometries\footnote{Euclidean wormholes are connected Euclidean spacetimes with multiple asymptotic regions.} correspond to closed big-bang / big-crunch cosmologies, as shown in Figure \ref{fig:MMcosmo}. Thus, a theoretical description of these Euclidean wormholes may lead to holographic models of cosmology.

For asymptotically AdS Euclidean wormholes with boundary geometries $B_i$, the AdS/CFT correspondence suggests that the physics of the wormhole should be related somehow to the physics of a set of Euclidean CFTs on $B_i$. The simple partition function for a set of uncoupled CFTs on disconnected geometries $B_i$ factorizes into the product of partition functions for the individual theories, so this partition function apparently does not carry information about the wormhole geometry.

\subsubsection*{Wormholes from ensembles}

It was suggested in \cite{Maldacena:2004rf} that a two-boundary wormhole might instead correspond to a sum
\be
\label{prodpart}
\sum_i Z_i^2 Z_i^2
\ee
of partition functions for a collection of CFTs.\footnote{This would be analogous to the way that Lorentzian wormholes correspond to a sum of product states $\sum_i c_i |\Psi_i \rangle \otimes |\Psi_i \rangle$ \cite{Maldacena:2001kr}.} Recently, this picture has been realized in the context of simple models of gravity, where the sum (\ref{prodpart}) represents an ensemble average over different boundary theories. For JT gravity in two dimensions, Euclidean wormholes with $n$ circular boundaries contribute to a gravitational path integral that corresponds to an ensemble average
\be
\label{partsum}
\langle Z_1(H) \cdots Z_n(H) \rangle_{p(H)}
\ee
where $Z_i$ is the thermal partition function of a quantum mechanical Hamiltonian $H$ and the average is over a certain probability distribution for possible Hamiltonians $H$  \cite{Saad:2019lba}. In the simplest possible topological theory of gravity in two dimensions, the gravitational path integral with $n$ circular boundaries has been shown to correspond to an ensemble average of partition functions as in (\ref{partsum}) where the Hamiltonians are trivial $H=0$ and the average is over the dimension of the Hilbert space \cite{Marolf:2020xie}. An example in the context of three-dimensional gravity has been described recently in \cite{Maloney:2020nni,Afkhami-Jeddi:2020ezh,Cotler:2020ugk}. Extensions of the JT-gravity story to asymptotically de Sitter spacetimes have been discussed in \cite{Maldacena:2019cbz,Cotler:2019dcj}.

\subsubsection*{Wormholes from interactions}

A more straightforward interpretation for the non-factorization of the partition function and correlators suggested by the connectedness of the wormhole is that the CFTs associated with various asymptotic regions are actually coupled somehow. If this is the case, they must be coupled in a way that the correlators between operators in different CFTs do not have short-distance singularities \cite{Betzios:2019rds}. Thus, we want the coupling to be very weak in the UV \cite{Betzios:2019rds}. On the other hand the coupling must be strong enough in the IR to give rise to a classically connected geometry in the gravity picture. It is a challenge to realize both of these features using conventional local interactions in quantum field theory.

\subsubsection*{Black Hole Microstate Cosmology as Maldacena-Maoz cosmology}

The goal of this note is to understand better how one can realize Euclidean wormhole solutions and thus models of closed universe cosmology using conventional holographic theories in higher dimensions.

\begin{figure}
\centering
\includegraphics[width=80mm]{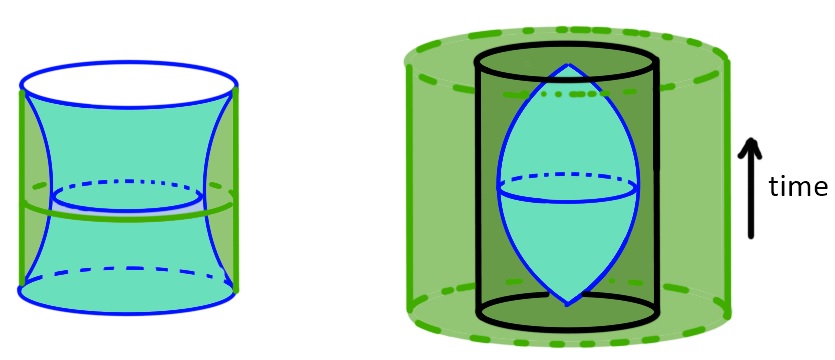}
\caption{The Black Hole Microstate Cosmology (BHMC) construction. Left: A holographic CFT theory on a Euclidean cylinder with boundary degrees of freedom at the two ends is dual to a Euclidean spacetime with an end-of-the-world (ETW) brane having the geometry of a Euclidean wormhole. Right: In the associated Lorentzian geometry, the ETW brane lives behind a black hole horizon and has the geometry of a closed-universe big-bang/big-crunch cosmology. With many boundary degrees of freedom in the original picture, gravity can be localized to this ETW brane.}
\label{fig:BHMC}
\end{figure}

We begin in section 2 by pointing out that a recently suggested model \cite{Cooper2018} for closed-universe cosmology within AdS/CFT can be understood as arising from a type of wormhole configuration in the Euclidean picture (Figure \ref{fig:BHMC}). We argue that the field theory description has aspects of both the ensemble picture and the interaction picture. We can understand the CFT description as starting from decoupled holographic theories, which are then coupled to some auxiliary degrees of freedom (Figure \ref{fig:auxiliary}). The effective coupling between the original theories is weak in the UV since the auxiliary degrees of freedom live on a higher-dimensional cylinder, and the original theories are physically separated by a Euclidean time direction, interacting with the fields at either end of the cylinder.

The partition function for the full system is equivalent to a type of ensemble average of partition functions for the original degrees of freedom,
\be
\label{ensemble_average}
Z = \langle Z[J_1] Z[J_2] \rangle_{p(J_1,J_1)}
\ee
where $Z[J_i]$ represents the original theory with certain sources turned on, and the average is defined by a joint probability distribution for the sources. The sources themselves arise as certain operators built from the auxiliary degrees of freedom, and the probability distribution arises from the path integral over the auxiliary degrees of freedom. In gauge theory models, we suggest that the sources should transform in a non-trivial representation of the gauge group (e.g. they should appear ``inside the trace'') in order to result in a classically connected wormhole.

\subsubsection*{Essential features for wormhole models of cosmology}

In section 3, we analyze the various elements appearing in the BHMC construction to understand which are essential and which can be generalized. We identify three key features that seem essential to obtain the right kind of interactions between the original CFTs to obtain a wormhole: (i) the existence of auxiliary degrees of freedom through which the original CFTs couple, (ii) coupling between the original degrees of freedom and the auxiliary degrees of freedom via non-singlet operators (i.e. coupling ``inside the trace''),\footnote{In the ensemble average picture, this implies that the relevant ensembles are over sources for the original theories that appear inside the trace.} and (iii) spreading out the auxiliary degrees of freedom spatially so that the original CFTs are physically separated.

We discuss ways to generalize the BHMC construction, explaining why this is difficult to do without the features listed above and how other features of the construction might be modified to obtain a broader class of cosmological models.

In section 4, we discuss a number of conceptual issues related to this approach to cosmology. We point out that properties (i) and (ii) together imply that the Euclidean path integral for the model can be associated to a specific pure state of the auxiliary degrees of freedom (Figure \ref{fig:WU}). This provides a ``wavefunction of the universe'' \cite{Hartle:1983ai}, which lives in the Hilbert space of the auxiliary degrees of freedom. In the BHMC example, these auxiliary degrees of freedom were a 3+1 dimensional CFT on $S^3$, but in the generalizations described in section 3, these can be as simple as a holographic matrix model or perhaps even a theory which is not conventionally holographic.

We also comment on the emergence of time and quantum mechanics, the relationship between the Euclidean and Lorentzian descriptions of the physics, and the compatibility of these time-symmetric universe models with physics in our own cosmological spacetime.

{\bf Note added:} While this manuscript was nearing completion, the papers \cite{Chen:2020tes,Hartman:2020khs} appeared, which discuss some related issues in models of cosmology. The paper \cite{Chen:2020tes} discusses lower-dimensional analogues of the BHMC construction (which was originally inspired by these lower dimensional constructions via \cite{Kourkoulou:2017zaj}).\footnote{See also \cite{Penington:2019kki, Dong:2020uxp, Chen:2020tes} for other discussions of cosmology in these low-dimensional models.} The work \cite{Hartman:2020khs} can be interpreted as providing evidence for the existence of auxiliary degrees of freedom in descriptions of closed-universe cosmology.

\section{Black Hole Microstate Cosmology and Euclidean wormholes}

In this section, we review the Black Hole Microstate Cosmology (BHMC) proposal of \cite{Cooper2018} for how to realize certain big bang/big crunch cosmologies within the context of AdS/CFT. We point out that this can be understood as a variant  of the Maldacena-Maoz picture where a cosmological spacetime arises from the analytic continuation of a Euclidean wormhole. We explain how the field theory construction can be thought of as arising from coupling a pair of Euclidean holographic CFTs to some auxiliary degrees of freedom or alternatively as a type of ensemble average for the original theories.

\subsection{Review of Black Hole Microstate Cosmology}

\begin{figure}
\centering
\includegraphics[width=70mm]{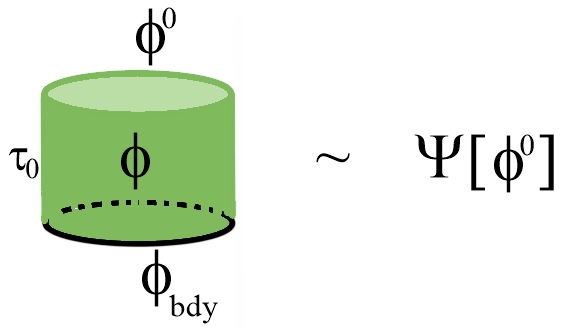}
\caption{Euclidean path integral construction for the state of a CFT on $S^d$ used in the BHMC construction. This is obtained from the standard path integral defining the vacuum state by truncating at some past Euclidean time $-\tau_0$ and choosing some boundary conditions there, with additional boundary degrees of freedom.}
\label{fig:PIstate}
\end{figure}

The basic setup in \cite{Cooper2018} considers states of a CFT on $S^d$ constructed using the Euclidean path integral for a BCFT shown in figure \ref{fig:PIstate}. This was inspired by the low-dimensional examples in \cite{Kourkoulou:2017zaj}. This is similar to the standard Euclidean path integral on $S^d \times [0, - \infty)$ used to construct the vacuum state of the CFT on $S^d$, except that we eliminate the region with Euclidean time $\tau < - \tau_0$ and place a boundary at $\tau = \tau_0$, possibly with some extra boundary degrees of freedom. For small enough $\tau_0$, the result is a pure high-energy state which has the interpretation on the gravity side as a pure black hole microstate. Similar black hole states were considered in \cite{Almheiri:2018ijj}.

The Euclidean path integral used to calculate properties of these states (associated with $\langle \Psi | \dots | \Psi \rangle$) ) has the geometry of a finite cylinder with boundaries at $\tau = \pm \tau_0$. For holographic theories, this is equivalent to the gravitational path integral for spacetimes whose boundary is the cylinder. This is expected to be dominated by a particular classical solution of the dual gravitational theory, and in \cite{Cooper2018} it was shown that in many cases, the dual solution has the topology shown in Figure \ref{fig:BHMC} (left), where the BCFT boundary extends into the bulk as an (end-of-the-world) brane that connects the two ends of the boundary cylinder. The Lorentzian geometry has an AdS-Schwarzschild exterior, but behind the horizon, we have an ETW brane that serves as an inner boundary of the spacetime. This emerges from the past singularity and terminates in the future singularity; its worldvolume geometry is that of a big-bang/big-crunch cosmology (Figure \ref{fig:BHMC}, right).

We see that the geometry of the ETW brane in this case takes the form of a Euclidean wormhole,\footnote{In fact, the whole geometry is topologically the same as a Euclidean wormhole, but the wormhole is ``fattened'' to include an asymptotically AdS region with the geometry of a cylinder.} while in the Lorentzian solution, the ETW brane geometry is that of closed universe cosmology. So we have a realization of the Maldacena-Maoz picture, but for the ETW brane. Generally, this doesn't provide a useful model of cosmology, since the gravitational physics is higher dimensional (i.e. not localized to the ETW brane). However, it was argued in \cite{Cooper2018} that when the number of boundary degrees of freedom is large relative to the number of local bulk degrees of freedom (corresponding to a large ETW brane tension on the gravity side), gravity can be localized to the ETW brane for a significant portion of the evolution (as in \cite{Karch:2000ct}).\footnote{This gravity localization was studied in more detail in \cite{Antonini2019} in the original setup and generalizations to charged black holes.} In these favorable situations, the BHMC setup seems to provide a specific realization of the Maldacena-Maoz approach to closed universe cosmology.

\subsection{BHMC as an interaction / ensemble average between decoupled theories}

In the introduction, we reviewed two previous suggestions for how to realize Euclidean wormholes within AdS/CFT. One was to consider initially decoupled Euclidean holographic theories and add interactions between them that are very weak in the UV (to avoid short-distance singularities in cross-correlators) and strong in the IR (to give a classically connected spacetime). Another was to consider an ensemble average of partition functions. In this section, we explain how the BHMC construction can be related to both of these approaches.

\subsubsection*{BHMC as an interaction between theories}

We recall that in order to obtain effective $d+1$-dimensional gravitational physics localized on the ETW brane starting from a $d+1$-dimensional cylinder, the analysis of \cite{Cooper2018} suggests that we should have many boundary degrees of freedom. A way to understand this is to think of the boundary degrees of freedom as separate $d$-dimensional holographic systems that are initially decoupled. On their own, these would be dual to $d+1$-dimensional gravitational theories. We end up with a wormhole by making these interact. Instead of introducing direct interactions between these theories, we couple them to auxiliary degrees of freedom, the bulk CFT degrees of freedom on the cylinder.\footnote{We will review in the next section why it is generally not sufficient to couple the theories directly.} Since the original theories couple to the two ends of the cylinder, there are no short-distance singularities in correlators of operators in the two original theories (now boundary operators on the two ends of the cylinder).

\begin{figure}
\centering
\includegraphics[width=120mm]{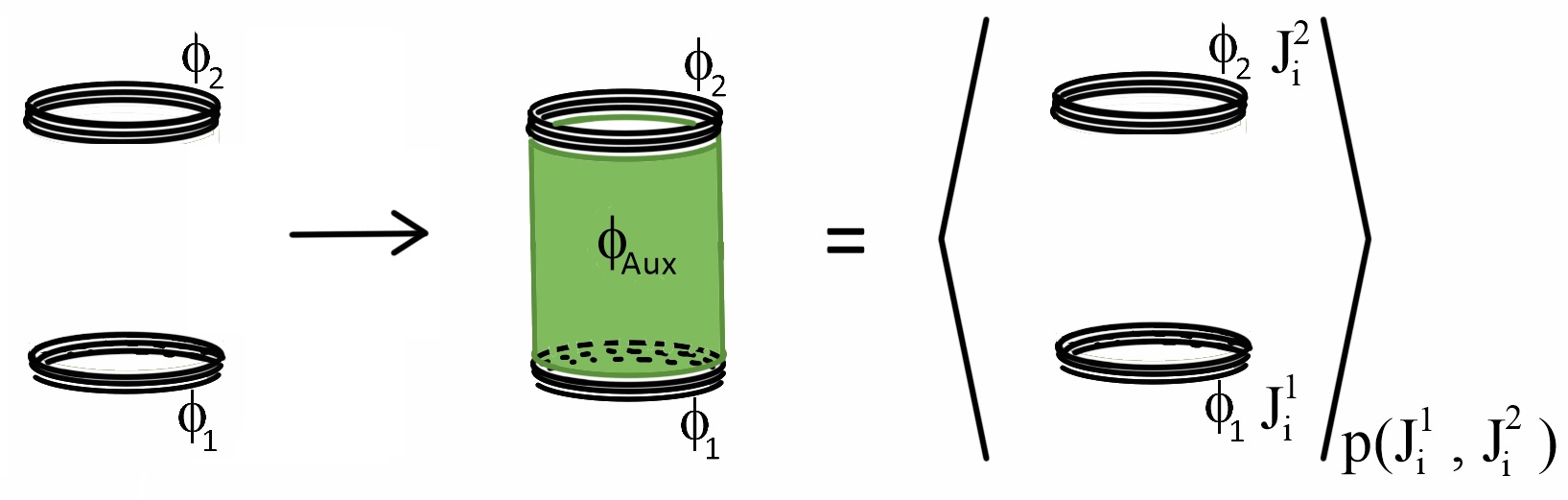}
\caption{Coupling the original CFTs to auxiliary degrees of freedom (a higher dimensional CFT on a cylinder in the BHMC case.) The partition function of the full theory is equal to the product of partition functions for the original theories averaged over an ensemble of sources defined using the path integral for the auxiliary degrees of freedom.}
\label{fig:auxiliary}
\end{figure}

If there are relatively few of these new degrees of freedom (e.g. the number of local bulk degrees of freedom on the cylinder is small compared with the number of local degrees of freedom in the original theory), it is plausible that some aspects of the dual gravitational physics is still $d+1$-dimensional\footnote{In the bottom-up effective description, we described this as gravity localization to an ETW brane, but more generally, the point is that we have some effective local $d+1$-dimensional description of the physics.}.  There seems to be a tradeoff here: the bulk cylinder degrees of freedom and their interactions with the boundary degrees of freedom need to be substantial enough to lead to a different topology for the ETW brane, but still weak enough so that the physics on large portions of the ETW brane is effectively $d+1$-dimensional.

\subsubsection*{BHMC as an ensemble average of theories}

The picture of two initially decoupled holographic theories interacting with some auxiliary degrees of freedom may also be understood as a type of ensemble average.

We can think of the action for the full system as
\be
S =  S_1 + S_2 + S_{Aux} + \int_{b_1} dx_1 {\cal O}_i(x_1) {\cal O}^1_i(x_1) + \int_{b_2} dx_2 {\cal O}_i(x_2) {\cal O}^2_i(x_2)
\ee
where ${\cal O}^{1,2}_i$ represent the local operators in the boundary systems and ${\cal O}_i$ are boundary operators in the higher-dimensional CFT.\footnote{To define the higher-dimensional CFT on the finite cylinder as an independent theory, we need to choose some boundary conditions at the two ends of the cylinder, so the auxiliary theory is some BCFT. We take this to have simple boundary conditions without additional degrees of freedom.}\footnote{As we discuss below, we should allow the index $i$ to include gauge indices (i.e. we should allow gauge-invariant interactions obtained by coupling non-singlet operators) in order to provide sufficiently strong interactions.}

We will now show that the partition function for the final higher-dimensional theory is equal to a certain ensemble average (\ref{ensemble_average}) of the partition functions for our original CFTs.

The full partition function is given by
\be
\label{part1}
\int [d \Phi] [d \phi_1] [d \phi_2] e^{ - S_1 - S_2 - S_{Aux} - \int_{b_1} dx_1 {\cal O}_i(x_1) {\cal O}^1_i(x_1) - \int_{b_2} dx_2 {\cal O}_i(x_2) {\cal O}^2_i(x_2)}
\ee
Now, we define the following joint probability distribution for sources $J^{1,2}_i$ in the two boundary theories:
\be
\label{prob}
p(\{J^1_i, J^2_i\}) \equiv \int [d \Phi]  e^{-S_{bulk}} \delta(J^1_i(x_1) - {\cal O}_i(x_1)) \delta(J^2_i(x_2) - {\cal O}_i(x_2))
\ee
Using this, we can rewrite (\ref{part1}) as
\beas
Z &=& \int [d J^1_i(x_1)][d J^2_i(x_2)] p(\{J^1_i, J^2_i\}) \int[d \phi_1] e^{- S_1 - \int_{b_1} dx_1 J_i(x_1) {\cal O}^1_i(x_1)} \int [d \phi_2] e^{- S_2 - \int_{b_2} dx_2 J_i(x_2) {\cal O}^2_i(x_2)} \cr
&=& \int [d J^1_i(x_1)][d J^2_i(x_2)] p(\{J^1_i, J^2_i\}) Z[ \{J^1_i\}] Z[ \{J^2_i\}] \cr
&=& \langle Z[ \{J^1_i\}] Z[ \{J^2_i\}] \rangle_{p(\{J^1_i, J^2_i\})}
\eeas
Thus, we see that the original partition function can be expressed as an ensemble average of the product of partition functions for the boundary theories with sources, where the average is taken with respect to a joint probability distribution for the sources generated by the bulk CFT path-integral. In the Euclidean context, adding sources for various operators amounts to changing the action, so we can think of each choice of sources as defining a different Euclidean field theory (typically with explicitly broken translation invariance). Thus, we can understand the result as a type of ensemble average, but the structure is somewhat more general since the probability distribution $p(\{J^1_i, J^2_i\})$ isn't diagonal (i.e. the two boundary theories are not always the same).

We recall from \cite{Cooper2018} that the connected ETW brane topology arises only for sufficiently small $\tau_0$.
From the ensemble average point of view, the effect of making $\tau_0$ (the cylinder height) small is to produce strong correlations between the sources in the distribution $p(\{J^1_i, J^2_i\})$. In the limit $\tau_0 \to 0$, we would recover the strict ensemble average picture where the sources are the same.

\section{Lessons and Generalizations}

In the previous section, we explained how the BHMC construction can be viewed as a type of wormhole in the Euclidean picture and thus an example of the Maldacena-Maoz appoach to cosmology. In this section, we try to understand which are the essential features of the BHMC construction and which features can be generalized, in order to understand more generally how to realize cosmological spacetimes using AdS/CFT.

\subsection{Auxiliary degrees of freedom}

We have described how the BHMC construction can be understood as starting from initially decoupled holographic theories (associated with the two ends of the cylinder) and coupling them via interactions with a higher-dimensional CFT on a cylinder.

The auxiliary degrees of freedom serve to provide an interaction between the original holographic theories that is weak in the UV, so that we have no short-distance singularities, but strong in the IR, so that the asymptotic regions will be connected. Could we achieve the same thing without auxiliary degrees of freedom? This was considered in some detail in \cite{Betzios:2019rds}; here, we briefly mention a few challenges with this approach.

The most direct way to couple the original CFTs is by adding a perturbation
\be
\label{direct}
\delta S = \lambda \int d^dx  {\cal O}_1(x) {\cal O}_2(x)
\ee
where the dimensions of ${\cal O}_1$ and ${\cal O}_2$ are chosen so that the interaction is relevant. For sufficiently low operator dimensions, this does give an interaction that avoids short-distance singularities \cite{Betzios:2019rds}. Specifically, we need
\be
\Delta_1 + \Delta_2 < {d \over 2}
\ee
so that $\langle {\cal O}_1(x_1) {\cal O}_2(x_2) \rangle$ has a finite limit as $x_1 \to x_2$. Assuming that $\Delta_1 = \Delta_2$ and that the operators satisfy the unitarity bound, we must have
\be
\label{dimbounds}
{1 \over 2} \le \Delta \le {3 \over 4}
\ee
in the case where we start with three-dimensional CFTs to model four-dimensional cosmology. Alternatively, \cite{Betzios:2019rds} suggested to consider non-local interactions
\be
\label{direct2}
\delta S = \lambda \int d^dx_1 d^d x_2 \lambda(x_1, x_2) {\cal O}_1(x_1) {\cal O}_2(x_2) \; .
\ee
These allow sufficiently soft UV behavior for a wider range of dimensions given an approriate choice of the source $\lambda(x_1, x_2)$.

Even if we obey these restrictions, a single coupling of the form (\ref{direct}) or (\ref{direct2}) where ${\cal O}_1$ and ${\cal O}_2$ are gauge-invariant operators is generally not strong enough in terms of $N$-scaling to reproduce the results from a classically connected wormhole configuration. Interactions of this type between two CFTs were considered in general in \cite{Aharony:2006hz} and more generally in \cite{Haehl:2019fjz} (see also \cite{Freivogel:2019lej}); it was found there that a perturbation $\lambda(x_1,x_2) {\cal O}_1(x_1) {\cal O}_2(x_2)$ does not modify the original disconnected background at the classical level unless we also have single-trace sources  $\lambda_1(x) {\cal O}_1(x)$ and/or $\lambda_2(x) {\cal O}_2(x)$ turned on. Even with these additional single-trace sources, the effects on the classical geometry were shown to be equivalent to turning on effective single-trace sources $\tilde{\lambda}_1(x)$ and $\tilde{\lambda}_2(x)$ without any interactions \cite{Haehl:2019fjz} (see also \cite{Witten:2001ua,Berkooz:2002ug}). These clearly cannot produce a connected geometry in the gravity picture.\footnote{These results were based on considering the effects of the added sources to all orders in perturbation theory. It would be interesting if non-perturbative effects could somehow modify these expectations.}

To overcome this issue with $N$-scaling, we could add many such interactions (as in the Lorentzian story of \cite{Maldacena:2018lmt}), but typical holographic CFTs generally do not have a sufficient number of operators of sufficiently low dimension. Even if they did, the corresponding gravity theory would have a very large number of light scalar fields, which is not desirable if we want to describe a somewhat realistic cosmology.

Thus, without introducing auxiliary degrees of freedom, it seems difficult to achieve interactions between the original CFTs that have the desired properties to produce a wormhole in the gravity picture.

\subsection{Interactions built from gauge non-singlet operators}

We have described in the previous section how interactions involving gauge-singlet operators in the original CFTs generally do not lead to a classically connected wormhole geometry. The BHMC example shows that introducing auxiliary degrees of freedom allows us to incorporate interactions between the original CFTs that are strong enough to produce a classical wormhole in the gravity picture. In this section, we will understand how this works more directly in a field theory construction.

A key point is that auxiliary matter can be taken to transform in some non-trivial representation of the gauge groups associated to the original theories. We can build singlet interactions from non-singlet operators in the original theory combined with non-singlet operators from the auxiliary degrees of freedom. For example, if the original theories are each $U(N)$ gauge theories, we can introduce bifundamental matter and add couplings such as $\tr(\Phi M_1 \Phi^\dagger M_2)$ or $\tr(\Phi M_1 \Phi^\dagger) + \tr(\Phi^\dagger M_2 \Phi)$ where $M_1$ and $M_2$ are fields or products of fields in the original CFT that transform in the adjoint representation of the respective gauge groups. In this case, the number of auxiliary degrees of freedom is similar to the number of degrees of freedom in the original theory.

If we would like the number of auxiliary fields to be parametrically smaller than the number of fields in our original theories, we can include an auxiliary theory with a third gauge group $U(n)$ and bifundamental matter between CFT1 and the new theory and between CFT2 and the new theory, with interactions $\tr(\Phi_{13}^\dagger M_1 \Phi_{13}) + \tr(\Phi_{23}^\dagger M_2 \Phi_{23})$. This has the advantage that we have a parameter $n$ that controls the amount of auxiliary matter, and thus the strength of the interaction between the original two theories.

\begin{figure}
\centering
\includegraphics[width=120mm]{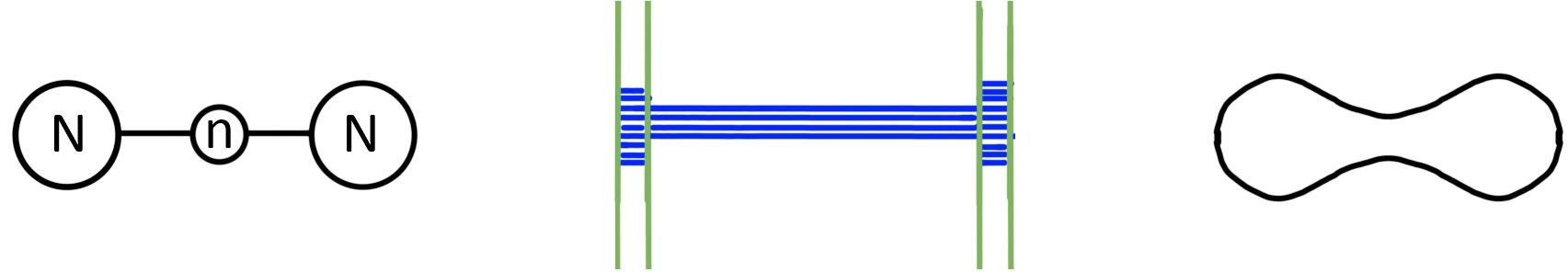}
\caption{Left: Schematic quiver diagram of two gauge initially decoupled gauge theory theories coupled to an auxiliary gauge theory via bifundamental matter. Center: D-brane construction of such a theory, with Dp-branes stretched between D(p+2)-branes and/or NS5 branes. The branes can be taken far apart at the same time as the decoupling limit so that the auxiliary degrees of freedom are spread over a finite extent in an extra dimension. Right: Geometry of the internal space for small $n/N$ in a conformal field theory example.}
\label{fig:quiver}
\end{figure}

We will describe in the next subsection how it is useful to distribute the auxiliary degrees of freedom spatially over an extra dimension. But consider first the simple situation where the auxiliary degrees of freedom are those of a field theory with the same dimension as the original theories and the interactions are local. Such field theories correspond to quiver gauge theories of the type shown in Figure \ref{fig:quiver}. In order to achieve interactions that are weak in the UV and strong in the IR, we should arrange that the couplings between the auxiliary matter and the original CFT are all relevant. For example, we can have $\tr(\Phi M_1 \Phi^\dagger M_2)$ where $M_i$ and $\Phi$ are scalar fields, in $d < 4$ or $d=4$ if the operator is marginally relevant. Or we can have $\tr(\Phi M_1 \Phi^\dagger) + \tr( \Phi^\dagger M_2 \Phi)$ in $d < 6$ or $d=6$ if the operator is marginally relevant. While these couplings will lead to interactions between the original theories that become stronger in the IR, the direct interaction examples above show that even with relevant interactions, we generally do not avoid short-distance singularities in correlators between operators in the two original CFTs. It would be interesting to investigate further to understand whether any models of this type have interactions that are sufficiently soft in the UV so that the gravity picture has two separate asymptotic regions.

Holographic quantum field theories similar to the ones we are describing, but with marginal interactions, were studied in \cite{Bachas:2017rch} and proposed as description of ``codimension 0'' wormholes. These theories are three-dimensional superconformal theories associated to quiver diagrams of the type shown on the left side of Figure \ref{fig:quiver}.\footnote{In general, the outside parts of the quiver can be more complicated, but the key point is to have relatively few degrees of freedom coupling the two sides.}  They arise from the low-energy description of D3-branes stretched between D5 and/or NS5 branes, as shown in Figure \ref{fig:quiver} (center). The geometries dual to the Lorentzian theories on $R^{2,1}$ are solutions of type IIB supergravity of the form
\be
ds^2 = f(\mu_i) ds^2_{AdS_4} + g_{ij} dx^i dx^j \; .
\ee
Thus, we have $AdS_4$ fibered over some compact internal space with metric $g_{ij}$. The geometry of this internal space depends on the ratio $n/N$ that controls the number of auxiliary degrees of freedom relative to the original. For small values of $n/N$, the internal space has a narrow throat connecting two larger regions, as shown in figure \ref{fig:quiver} (right). Thus, the interactions provided by the auxiliary $U(n)$ gauge theory give rise to a bridge between two otherwise-disconnected regions of the internal space.\footnote{The full AdS$_4$ is fibered over this internal-space wormhole, so the wormhole is codimension 0 from the point of view of the non-compact directions of spacetime.} Thus, as promised, auxiliary matter with non-singlet interactions has a geometrical effect, giving rise to a bridge between two otherwise-disconnected regions of the internal space.

It is possible that a wormhole geometry could be obtained by considering a non-conformal version of this where this bridge varies as a function of the radial direction, from a disconnected space in the UV to a single connected internal space in the IR. In field theory language, this suggests that the effective number of auxiliary degrees of freedom, or the strength of their interaction, should increase going from UV to IR, as we would have if the interactions with the auxiliary degrees of freedom are relevant. However, it is essential that the interactions that are weak enough in the UV to correspond to a disconnected internal space. Next, we discuss the feature of the BHMC model that achieves this UV softness.

\subsection{Spatially distributed auxiliary degrees of freedom}

Incorporating auxiliary degrees of freedom that have non-singlet interactions with the original theories allows interactions that are strong enough in the infrared to result in a geometrically connected spacetime for the gravitational dual. It is still a challenge to simultaneously obtain interactions that are weak enough in the UV so that we have no short-distance singularities, or equivalently, that we end up with two distinct asymptotic regions in our geometry.

The feature of the BHMC model that achieves this UV softness is that the auxiliary degrees of freedom are spread over an extra spatial direction, so that the original CFTs are physically separated. In this case, correlators involving an operator in one CFT and an operator in the other CFT at the same point on the sphere are still separated by the extra dimension, so we naturally avoid any short-distance singularities.

We could also consider auxiliary degrees of freedom which live on some more general higher-dimensional geometry. A very simple generalization is to ``double'' the cylinder so that we auxiliary degrees of freedom on $S^d \times S^1$, with the original theories living on opposite points of the $S^1$. This gives rise to a two-sided version of the BHMC story, where the RS-brane is a defect brane rather than an ETW brane, so we would have a more conventional Randall-Sundrum cosmology \cite{Randall:1999ee, Randall:1999vf,Karch:2000ct}. The Euclidean version of this setup has been considered recently in \cite{Chen:2020uac} (see figure 2 of that paper).

\subsection{Generalizing the original and auxiliary CFTs}

In the BHMC construction, we took the original theories associated with the ends of the cylinder to be conformal theories on $S^d$, but we could just as well consider non-conformal theories, or generalize to other geoemtries such as flat Euclidean space. This would provide more possibilities for the spatial geometry of the universe and the evolution of the scale factor in the cosmological description.

We can also consider generalizing the auxiliary degrees of freedom. The Euclidean geometry associated with BHMC has the topology of a Euclidean wormhole, but the wormhole is ``fattened'' to include an asymptotically AdS cylinder. This aspect of the geometry is associated with the fact that our auxiliary degrees of freedom take the form of a $d+1$-dimensional CFT. The most obvious way to eliminate the asymptotically AdS feature  (and thus obtain a more conventional closed universe cosmology rather than a Randall-Sundrum cosmology) would be to eliminate the UV degrees of freedom in the $d+1$-dimensional cylinder CFT using some type of cutoff. There are some interesting possibilities for how to introduce a cutoff while maintaining some control over the holographic description.

\subsubsection*{Approximating the $S^d$ field theory by a matrix model.}

One way to cut off a quantum field theory on $S^d$ is by truncation to a finite set of spherical harmonics for the fields. This gives rise to a non-commutative version of the theory, which can often be understood as arising from a 0+1 dimensional matrix model. For example, the ${\cal N}=4$ SYM theory on $S^3$ arises from the large $N$ limit of a supersymmetric matrix model (known as the BMN matrix model or Plane Wave Matrix model \cite{Berenstein:2002jq}) expanded about a particular vacuum \cite{Kim:2003rza}.

In passing to a cutoff version of the bulk theory on the cylinder, it may be natural to replace the original CFTs on $S^d$ with regulated versions. This should truncate the asymptotically AdS boundaries in the Euclidean wormhole geometry, and lead to some softening of the big-bang / big-crunch singularities in the Lorentzian picture.\footnote{We note that such softening is not necessary, since the original theory with continuum CFTs is still well-defined and should precisely encode the physics of the singularities in the gravitational picture.} The final Euclidean theory would be a matrix model living on an interval $[-\tau_0, \tau_0]$ of Euclidean time with additional matrix degrees of freedom living at the two ends of the interval.\footnote{There may be natural models of this type arising in string theory from Euclidean D0-branes stretched between D2-branes and/or NS5-branes as in Figure \ref{fig:quiver} (center).} The associated path integral on the interval $[-\tau_0, 0]$ produces a particular quantum state of the BMN matrix model.

\subsubsection*{$T\bar{T}$ deformations and UV-free theories}

There are various interesting alternatives to cutting off the cylinder CFT. We could consider a $T\bar{T}$-deformation \cite{Zamolodchikov:2004ce,Smirnov:2016lqw,Cavaglia:2016oda} of the CFT on the cylinder, which has been argued \cite{McGough:2016lol} to correspond to moving the asymptotic boundary of AdS to a finite radius. $T\bar{T}$-deformations have also found an application in dS/dS models of de Sitter space \cite{Gorbenko:2018oov}.

Instead of introducing a cutoff, we could replace the degrees of freedom on the cylinder with some non-conformal theory that is free or weakly coupled in the UV but flows to a holographic CFT in the IR, or even a theory that is not holographic. In this case, the part of the geometry with a geometrical description would not include an asymptotically AdS region and in some cases, we might avoid a geometrical black hole exterior region all together.

A final idea for a simplification would be to replace the Euclidean time direction of the cylinder CFT with a discretized version, or alternatively a ``deconstructed'' version using the ``Deconstructing Dimensions'' idea of \cite{ArkaniHamed:2001ca}. Here, the Euclidean time direction of the cylinder emerges only in the IR, starting from a particular quiver gauge theory.

\subsection{Lessons for ensemble averages}

To conclude this section, we decribe briefly how the various key features of the BHMC construction manifest themselves in the ensemble average picture.

Our first essential ingredient, the existence of auxiliary degrees of freedom, is what gives rise to an ensemble average picture in the first place. As we have seen in section 2, the auxiliary operators coupling to the original CFTs can be reinterpreted as sources, and the path integral over the auxiliary degrees of freedom can be interpreted as generating a joint probability distribution for the sources.

The second essential ingredient, the non-singlet nature of the interactions with the auxiliary degrees of freedom, corresponds to having an ensemble average of theories defined by adding sources to the original theory, where the sources couple to gauge non-singlet operators. In simple cases where the fields in the original theories transform in the adjoint representation of a $U(N)$ gauge group, we would have sources that also transform in the adjoint representation, appearing inside the trace in the action.

The final ingredient, that the auxiliary degrees of freedom are spread out spatially over an extra Euclidean time direction, will lead to a joint probability distribution for the sources in which the sources for operators in the two original theories at equivalent points on the sphere are not too strongly correlated.

In the end, the full picture involving interactions with auxiliary degrees of freedom seems to give a more complete description, but we point out the ensemble average picture in order to make contact with the lower dimensional examples and because the ensemble-average picture may provide a more efficient description of some aspects of the physics for which the precise nature of the auxiliary degrees of freedom are unimportant.

\section{Discussion and Questions}

In this paper, we have understood that the BHMC construction provides a particular realization of the Maldacena-Maoz picture of obtaining cosmological physics from a Euclidean wormhole, and suggested various ways to generalize this. In this final section we point out a few interesting features of this approach to cosmology.

\subsection{A quantum state of the universe}

We have seen that two key features of the BHMC construction are that we have some auxiliary degrees of freedom and that these degrees of freedom are distributed over an extra Euclidean time direction.

\begin{figure}
\centering
\includegraphics[width=80mm]{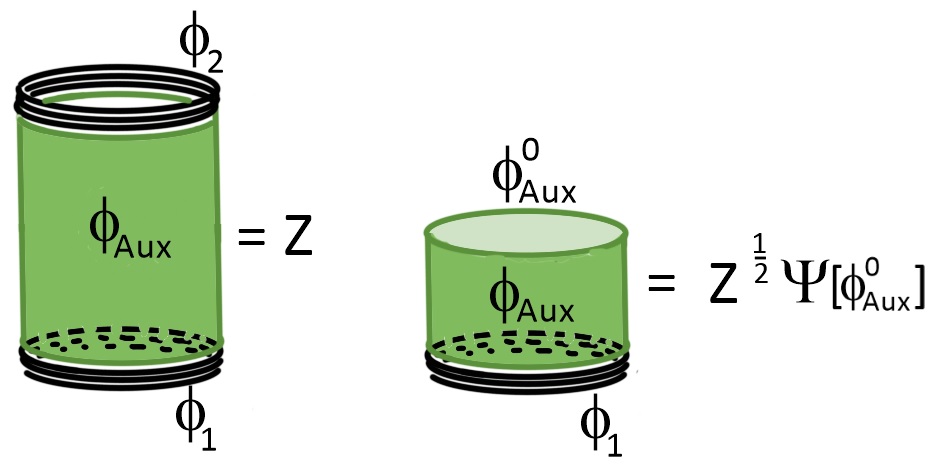}
\caption{Left: Euclidean path integral defining the model. Right: Euclidean path integral defining a state of the auxiliary degrees of freedom. This gives the wavefunction of the universe, encoding the Lorentzian cosmological spacetime.}
\label{fig:WU}
\end{figure}

A very interesting consequence of these features is that the full path integral for the theory,
\be
Z = \int_{\tau \in [-\tau_0, \tau_0]} [d \phi_1] [d \phi_2] [d \phi_{Aux}(\tau)] e^{-S_1 - S_2 - S_{Aux} - S_{int}}
\ee
can be associated with a quantum state in a Hilbert space associated with the auxiliary degrees of freedom. Specifically, we have the wave-functional
\be
\Psi[\phi_{Aux}] = {1 \over Z^{1 \over 2}}\int^{\phi_{Aux}(0) = \phi_{Aux}}_{\tau < 0} [d \phi_1] [d \phi_{Aux}(\tau)] e^{-S_1 - S_{Aux} - S_{int}} \; .
\ee
In the BHMC example, this is the wavefunction for our bulk CFT on $S^d$ times time, associated to a black hole microstate of the dual Lorentzian gravitational theory. In a similar way, we can construct a quantum state in any construction with auxiliary degrees of freedom spread over an interval in Euclidean time, as shown in Figure \ref{fig:WU}.\footnote{A discretized version of the Euclidean time, or the version appearing in \cite{ArkaniHamed:2001ca}, may also be sufficient here.}

Thus, two of the features that allowed interactions with the right properties to lead to a wormhole in the Euclidean gravity picture combine to imply that we can associate some quantum state to our Lorentzian spacetime, i.e. that there is a well-defined wave function of the universe living in the Hilbert space of the auxiliary degrees of freedom.

An interesting point is that it may not be particularly important which degrees of freedom are chosen to host the wavefunction of the universe. As described recently in \cite{VanRaamsdonk:2018zws,Simidzija:2020ukv}, starting from Euclidean path integral states of the type we are considering here, there are natural maps to states in the Hilbert spaces of other CFTs or collections of BCFTs that give states for which the asymptotic region is different, but the interior of the region spacelike separated from the boundary time slice where the state lives is almost unaffected. This region can contain the entire cosmology.

The description of a cosmological spacetime via the auxiliary degrees of freedom is very analogous to the recently discussed description of black hole interiors via the state of radiation degrees of freedom in models of black holes \cite{Maldacena:2013xja, VanRaamsdonk:2013sza, Penington:2019npb, Almheiri:2019psf, Almheiri:2019hni}. This connection was emphasized in \cite{Sully:2020pza}. In the evaporating black hole context, we have something parallel to the discussion in the previous paragraph, where the encoding of the black hole interior by auxiliary degrees of freedom does not seem to care precisely what these auxiliary degrees of freedom are.

Another connection with the evaporating black hole models is that the wormholes appearing in the calculation of Renyi entropies for the radiation degrees of freedom \cite{Almheiri:2019qdq, Penington:2019kki, Rozali:2019day} appear in a very similar way to what we have described in this paper: we have multiple copies of some original quantum system (the degrees of freedom describing the black hole), and these are coupled to auxiliary degrees of freedom which are distributed on a higher-dimensional space (the auxiliary radition system living on the replica manifold). From the point of view of the original CFTs, the full path integral including the auxiliary degrees of freedom corresponds to an ensemble average of the original CFTs.

\subsubsection{Lorentzian vs Euclidean}

It is interesting to ask whether the wavefunction of the universe as we have described it always contains complete information about the dual Lorentzian geometry (i.e. about the cosmological spacetime). From the point of view of the original path integral, observables in the quantum state $|\Psi \rangle$ correspond to insertions of operators at $\tau = 0$ or on some Schwinger-Keldysh contour inserted at $\tau = 0$. But there are more general observables that we can calculate from the full path-intergal, such as correlators between operators in the original theories associated with the asymptotic regions of the wormhole, or correlators involving both the original and the auxiliary degrees of freedom.

Certain cosmological observables (e.g. correlators between fields on the ETW brane at $t=0$) seem relatively simple to compute using the full path integral (e.g. via an HKLL construction in the Euclidean picture), but much more difficult to compute directly from the state $|\Psi \rangle$, where they involve local physics behind the horizon of the black hole. But if $|\Psi \rangle$ has complete information about the Lorentzian spacetime, there should be a way to go between the Euclidean and Lorentzian pictures. It would be interesting to understand this better.

\subsection{Emergence of time and quantum mechanics}

Even when the physics of the cosmological spacetime is completely encoded in a quantum state $|\Psi \rangle$ of the auxiliary degrees of freedom, the time evolution in the corresponding quantum system has little to do with the cosmological time evolution. Indeed, it is expected that the state $|\Psi (t=0)\rangle$ encodes the geometry in the Wheeler-deWitt patch associated with the $t=0$ slice on the boundary (i.e. the region of spacetime spacelike separated from this boundary slice), and this can include the full cosmological evolution of the ETW brane.

Thus, it is interesting to understand how time evolution and quantum mechanics in the effective field theory description of the cosmology emerges either from the physics of the Lorentzian state $|\Psi \rangle$ or from the point of view of the original Euclidean path integral.\footnote{Even for the vacuum state of a holographic CFT, it is interesting to understand how time evolution in the Wheeler-de-Witt patch can be understood to emerge from properties of the CFT state.}

\subsubsection*{Time-symmetric cosmology}

An interesting feature of cosmological scenarios arising from analytically continued Euclidean wormholes is that the resulting universe (in the simplest cases) is symmetric under time-reversal (or more generally, CPT), i.e. there is no distinction between the future and the past in the complete description.\footnote{We could consider more a more general Euclidean setup which is not symmetry (e.g. by introducing complex sources), but in this section, we will explore the question of whether it is necessary to break the time symmetry in the fundamental description.} This may seem to contradict our expectations for big-bang cosmology, where an initially homogeneous universe develops structure, etc...

We recall that the conventional understanding is that the arrow of time is determined by the direction of increasing entropy: for some reason, the big bang was a time of low entropy, and the increase in entropy moving away from the big bang leads to our identification of the big bang as being in the past and the time direction moving away from the big bang is the forward direction. In a time-symmetric picture, there is no fundamental distinction between the big bang and the big crunch, so it is clearly not correct to say that the big bang was characterized by low entropy.

These issues have been discussed long ago in \cite{Page:1985ei} and various later papers. The basic point is that while the full description may be time-symmetric, this can simultaneously describe various branches of the wavefunction, some of which have forward thermodynamic evolution and some of which have backward thermodynamic evolution. A simple toy model is to consider a quantum state $|\Psi_+ \rangle$ describing the evolution of a system with increasing coarse-grained entropy increases during a time interval $[-T,T]$ superposed with the time-reversed state $|\Psi_- \rangle$ for which the arrow of time is reversed. In this example, a time-symmetric state in a conventional quantum system simultaneously describes two classical pictures with opposite directions of time.\footnote{As we have emphasized in the previous section, in the fundamental description, the cosmological evolution is more naturally associated with a quantum state at a particular time (as opposed to the full unitary evolution of this state), so our simple quantum mechanics example may be missing elements of the full story.} We could have a similar story in the time-symmetric cosmology, where the full picture simultaneously describes many possible semiclassical universes, some of which have one entropic direction of time and some of which have the other entropic direction of time. These various semiclassical pictures (including also the ensemble of possible realizations of structure) should arise from the complete gravitational wavefunction via a decoherence picture (e.g. tracing over the bulk degrees of freedom away from the ETW brane).

\section*{Acknowledgements}

We would like to thank Stefano Antonini, Alex May, Shiraz Minwalla, Moshe Rozali, Douglas Stanford, and Brian Swingle for useful discussions and comments. This work is supported by the Simons Foundation via the It From Qubit Collaboration and a Simons Investigator Award and by the Natural Sciences and Engineering Research Council of Canada.

\bibliographystyle{jhep}
\bibliography{references}

\end{document}